# Vehicle Class, Speed, and Roadway Geometry Based Driver Behavior Identification and Classification

Awad Abdelhalim, Montasir Abbas
Via Department of Civil and Environmental Engineering
Virginia Polytechnic Institute and State University
Blacksburg, Virginia, 24061
atarig@vt.edu, abbas@vt.edu

*Abstract*— Over the past decades, intense emphasis has been placed on understanding car-following behavior and the factors that affect it. The car-following process, however, still remains a very complex field of study in spite of all the efforts. This paper focuses on the study of the effect that the class of the vehicle, leading heavy vehicles in particular, causes on the following vehicle's behavior, specifically in terms of the following distance (gap) that the following vehicle keeps from the lead vehicle. This was done by extracting and analyzing different car following episodes in the Next Generation Simulation (NGSIM) dataset for Interstate 80 (I 80) in Emeryville, California, USA. The results of the statistical analysis are compared to that of the synthesized literature of research efforts that have been carried out on the topic, then are further assessed utilizing different calibrated behavioral clusters for the Gazis-Herman-Rothery (GHR) car-following model to address the similarities and differences in car following behavior between drivers of the same vehicle class. The paper also validates the results of the statistical analysis and highlights possible future implementations.

## I. Introduction

The car-following process is generally defined as the interaction between two consecutive vehicles in a traffic stream. The following car in a car-following episode is typically affected by the movement of the lead vehicle, accelerating and decelerating in order to maintain a specific desired spacing (gap) from the lead vehicle. This desired distance is mainly a following driver's preference, which is also affected by various factors, including the driver's psychology, perception and reaction time, traffic stream characteristics, vehicle type and condition, etc.

Most of the literature on the topic and the various representative models relates the car following spacing preference to the following driver's reaction time in addition to characteristic-specific parameters, which in itself are defined and affected by the driver's characteristics including age, sex and mental condition and so on. In this paper, however, the main factor to be assessed is the class (size) of the lead vehicle, and how it affects the gap that the following driver would typically prefer to keep ahead. The study was conducted for the NGSIM dataset for a segment of I-80 in Emeryville, California, containing trajectories of vehicles in a heterogeneous traffic stream in a 6 - lane freeway section. In the process, several assumptions and propositions have been made to classify and refine the trajectories of the vehicles in the dataset.

The final outcomes of the paper are compared to those perceived by the authors from the literature synthesized, before utilizing the outcomes of the Two-Step Clustering Algorithm by Higgs and Abbas [1] [2] to further assess the behavioral differences between drivers of the same vehicle class, ultimately providing suggestions for the implications of the results and recommendations for further possible future work on the topic.

### A. Objective

The major objective of this study is to study the car following behavior of passenger car drivers in a heterogeneous traffic stream. This is achieved by tracking multiple car-following episodes from the NGSIM trajectories dataset and retrieving the data for following distance between lead and following vehicle according to the lead and following vehicle's respective classes and the average speed observed during the car-following episode.

### B. NGSIM Dataset

The dataset of vehicle trajectory data utilized for this study is the Federal Highway Administration's (FHWA) Next Generation Simulation (NGSIM) project dataset for I-80. The data analyzed in this paper represent vehicle trajectories on a segment of I-80 collected between at the time between 4:00 - 4:15 P.M on April 13, 2005. The NGSIM dataset also includes data collected between time periods 5:00 - 5:15 P.M, and 5:15 - 5:30 P.M, but as those are mainly data of the congested section until it clears up and for the purpose of this study, varying following speeds is a parameter to be considered, therefore, only data for the first time span 4:00 - 4:15 P.M was analyzed, which included a total of 2911 vehicles passing through the monitored segment.

"This data was collected using video cameras mounted on a 30- story building, Pacific Park Plaza, which is located in 6363 Christie Avenue and is adjacent to the interstate freeway I-80. The University of California at Berkeley maintains traffic surveillance capabilities at the building and the segment is known as the Berkeley Highway Laboratory (BHL) site. Figure 1 provides a schematic illustration of the location for the vehicle trajectory dataset. Video data were collected using seven video cameras. Digital video images were collected over an approximate 5-hour period from 2:00 p.m. to 7:00 p.m. on April 13, 2005. Complete vehicle trajectories were transcribed at a resolution of 10 frames per second." [3]

The NGSIM dataset, however, has some level of unreliability in the trajectories recorded, Montanino and Punzo [4], and Punzo, Vincenzo, Borzacchiello, and Ciuffo [5] [6] have comprehensively studied the dataset and discussed techniques and methods to improve and refine the trajectories.

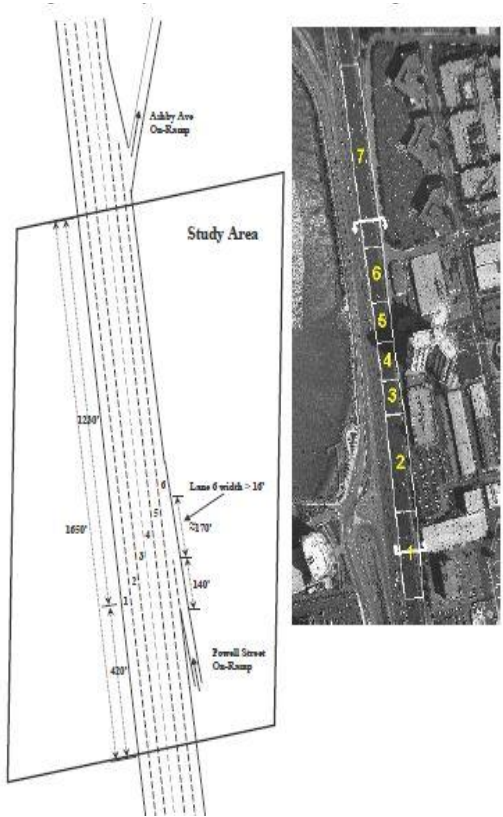

Fig. 1. Segment location at interstate 80 segment in Emeryville, California [3].

## II. LITERATURE SYNTHESIS

Evans and Wasielewski studied the "Risky Driving Related to Driver and Vehicle Characteristics" [7], where emphasis was placed on the factors that affect the tendency of a following vehicle driver to take more risks during a car following episode. The authors used data from video tapes taken by cameras on equipped vehicles alongside road camera records from Michigan, USA and Toronto, Canada. The data collected was statistically analyzed to find the variance and standard deviation with respect to multiple factors that include the driver's age, sex, road condition, time, lead vehicle class, follower vehicle's type, following driver's psychological behavior etc. It was concluded that more risks were associated with male drivers over female drivers, younger drivers over older drivers, drivers who drive alone over those who were accompanied and drivers who drive newer vehicles. One of the notices as well was that the drivers who do not wear seat belts, as a sign of psychological risk acceptance, would also have overall riskier driving, where they maintain relatively smaller gaps with the lead vehicle.

Green and Yoo [8] studied the "Driver Behavior While Following Cars, Trucks, and Buses", where the authors had participants use a driving simulator to study the preferred spacing they would keep when the front vehicle is a passenger car, a pickup truck, a school bus and a tractor trailer, while grouping the participants into four age/gender groups of young and old men and women. Participants were instructed to only follow the lead vehicle without attempting to overpass it, with no vehicles besides or behind the simulator vehicle, which adds a lot of uncertainty to the research outcomes. The research however concluded that participants followed passenger car at distances 10% closer than they followed larger vehicles.

Sayer, Mefford and Huang [9] analyzed data obtained from research vehicles equipped with video cameras and Intelligent Cruise Control (ICC) to calibrate headway time margin with respect to age and sex of the driver, grouping male and female drivers into groups of young, middle age and old drivers. The author's observations and analysis concluded that the passenger car drivers, on average, follow light trucks at shorter distances gaps and headway time margins than they follow other passenger cars, with some variance between gender and age groups. The authors explained this non-intuitive conclusion as a result of multiple factors that could include the state of traffic stream and the psychological effect it causes on the following driver's behavior.

In a more comprehensive recent study, Aghabayk, Sarvi, and Young [10] studied the car following process under different pairs of passenger car-heavy vehicle combinations as lead and following vehicles, subcategorized into passenger car following passenger car, passenger car following heavy vehicle, heavy vehicle following passenger car and heavy vehicle following heavy vehicle. The authors investigated the car following behavior of those lead/follower pairs in congested heterogeneous traffic condition in the NGSIM dataset for I-80, testing how different types of vehicles interact, and the resulting time and space headways, driver reaction time, vehicle's speeds and accelerations, introducing different Wiedemann model thresholds for each leader/follower combination. The research concluded that highest following headways would be in the heavy vehicle following heavy vehicle pair, the shortest are associated with a passenger car following another passenger car, while the way passenger and heavy vehicles follow one another would vary with respect to the overall speed. It was also noted that the reaction time for both passenger car and heavy vehicle's drivers increases 10% more when following a heavy vehicle than a passenger car, which indicates more difficulty in decision making process, while heavy vehicle drivers generally have more steady following behavior than that of passenger car's drivers.

This study has successfully identified and succeeded to explain many of the characteristics of car following behavior in heterogeneous traffic. However, only the analysis of headways between follower/leader pairs does not suffice for the complete understanding of the subject, as heavy vehicles are generally much longer and therefore are typically associated with larger space headways.

Higgs and Abbas [1] have proposed an algorithm to classify and group the driving behavior into certain structures based on driving patterns and states observed in car following episodes obtained from naturalistic driving data. A two-step segmentation and clustering algorithm was used to classify drivers of passenger cars and trucks into 30 discrete state-action clusters for each vehicle class drivers. The study concluded that car drivers showed discrete following behavior which is not well represented by only 30 behavioral clusters, while truck drivers were noticed to show common driving pattern regardless, implying that they take similar decisions and actions during the car-following process, and could be well represented by few behavioral clusters.

## III. METHODOLOGY

### A. Dataset Statistical Analysis

To overcome what seems to be weaknesses in the previously mentioned studies, this study analyzed the NGSIM's I-80 freeway section in the "build up to congestion" period, which would give the opportunity to study the car following behavior under various speeds conditions, and compare the findings in terms of following distance from the front bumper of the following vehicle to the rear bumper of the lead vehicle (gap) as opposed to the space headways concluded in the previously mentioned literature.

The prodigious dataset was initially subject to a statistical analysis of the data. Certain assumptions had to be taken in order to refine and classify the data before analysis of the car following episodes, these assumptions were:

- Passenger cars are identified as those vehicles with maximum length of 5m (16.5 ft.) or below.

- Heavy vehicles are identified as the vehicles with length greater than 5.5m (18ft.), which is including larger pick-up trucks.

- Vehicles of length 5-5.5m (16.5-18 ft.) are identified as SUVs and light pickup trucks.

- Vehicles must enter area in following episode, but not necessarily be in the episode all the way to the end of section. This is to assure that the interaction with the vehicles entering the monitored segment at the merge section is not mistaken for a car-following process.

- Few cases of following episodes with headways resulting in average overall gaps less than 4.5m (15 ft.) and greater than 76m (250 ft.) were eliminated, as they present NGSIM dataset trajectory errors, extreme aggressiveness or minimum to no influence.

- Vehicle trajectory is taken in Y direction co-ordinates included within the dataset.

- Average speeds and corresponding space headways are taken at varying speeds for each follower/leader pair, from which the average gap is taken as the difference between the average space headway from analyzed data and the length of the lead vehicle.

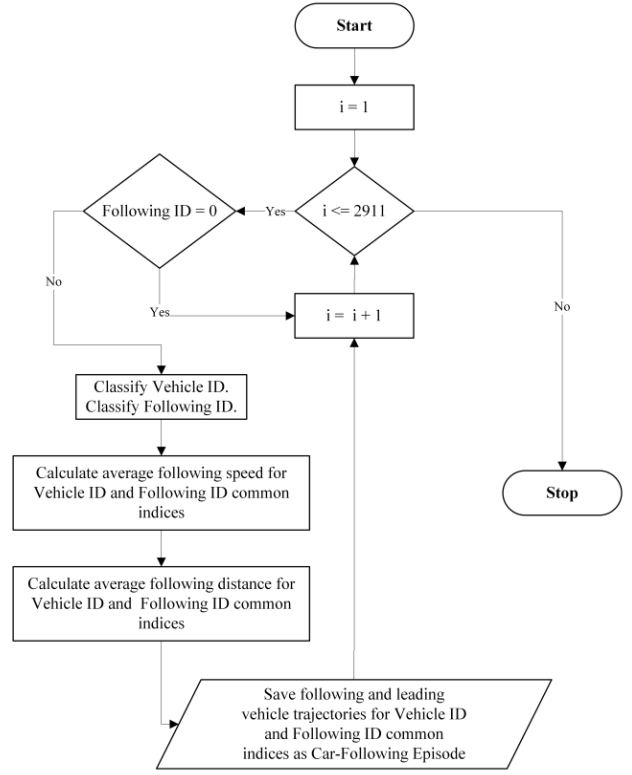

Fig. 2. Flowchart of the data flow for primary analysis.

### B. Extraction and Categorization of Car-Following Episodes Followed by the Application and Regression of Predefined Behavioral Cluster

To further assess and validate the results obtained from statistical analysis of the dataset trajectories, a MATLAB [11] code was written to extract and classify the car following episodes obtained from the dataset, and then use the clusters of predefined parameters for the Gazis–Herman–Rothery (GHR) model by Higgs and Abbas [1] to find the best-fitting behavioral cluster for each of the following drivers. The GHR model predicts the acceleration of the following vehicle through the following equation:

$$a_n(t) = c v_n^m(t) \frac{\Delta v(t-\tau)}{\Delta x^l(t-\tau)} \quad (1)$$

Where:

n = Index of the following vehicle.

$a_n(t)$ = Acceleration of the subject vehicle at time t.

$v_n(t)$ = Speed of the subject vehicle at time t.

τ = Perception reaction time of the subject vehicle's driver.

Δv (t − τ) = Speed of the subject vehicle at time (t − τ).

Δx (t − τ) = Space headway of the subject vehicle at time (t − τ).

c, l, m = GHR model parameters.

Tables VI and VIII in the paper by Higgs and Abbas [1] show the calibrated GHR parameters for 30 car and truck driver behavioral clusters, respectively. Each of the 1232 car-following episodes (that lasted a minimum of 25 seconds) obtained from the analysis of the NGSIM dataset were tested against 30 clusters based on the class of the following vehicle, and the cluster that produces the minimum root mean squared error was taken as a representative of that specific episode.

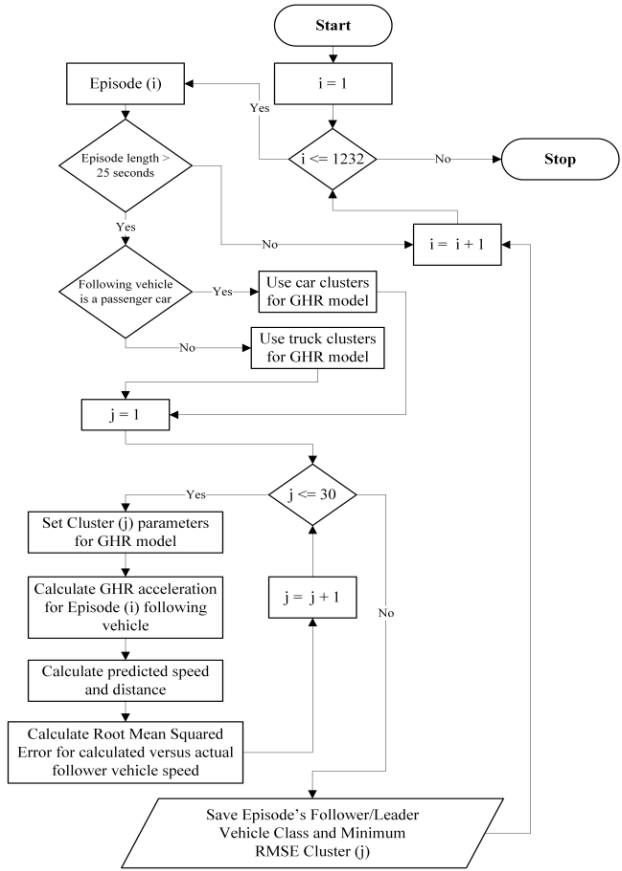

Fig. 3. Flowchart showing the data flow of the MATLAB code used for GHR regression.

## IV. RESULTS

### A. Dataset Statistical Analysis

The statistical analysis carried out under the mentioned conditions for 2911 vehicles has resulted in 1232 car-following episodes. Table 1 below shows the number of car-following episodes observed for each follower/leader vehicle class pairs and the respective average following distance (gap) and following speeds observed.

TABLE I. AVERAGE GAP AND FOLLOWING SPEED OBSERVED FOR DIFFERENT FOLLOWER/LEADER PAIRS.

| Passenger Car Following Heavy Vehicle | | | Passenger Car Following Passenger Car | | |
|---|---|---|---|---|---|
| Number Observed | Avg. Gap (m) | Avg. Speed (km/hr) | Number Observed | Avg. Gap (m) | Avg. Speed (km/hr) |
| 118 | 15.5 | 30 | 982 | 16.7 | 28.5 |
| Heavy Vehicle Following Passenger Car | | | Heavy Vehicle Following Heavy Vehicle | | |
| Number Observed | Avg. Gap (m) | Avg. Speed (km/hr) | Number Observed | Avg. Gap (m) | Avg. Speed (km/hr) |
| 114 | 30.7 | 32 | 18 | 37.7 | 29.8 |

It was observed that on average statistics for the analyzed episodes, passenger cars followed heavy vehicles almost 7% closer than they followed another passenger car. To be able to analyze the cause of this outcome, the analyzed episodes were broken down according to the following speed range. Table 2 below shows a comparison of the percentage of decrease in the gap for passenger cars following heavy vehicles versus following other passenger cars.

TABLE II. AVERAGE GAP DIFFERENCE OBSERVED FOR PASSENGER CARS CAR-FOLLOWING EPISODES.

| Passenger Car Following Heavy Vehicle | | Passenger Car Following Passenger Car | | Gap Decrease % |
|---|---|---|---|---|
| Speed < 32.2 km/hr (20 mph) | | | | |
| Number | Avg. Gap (m) | Number | Avg. Gap (m) | 13.25 |
| 76 | 12 | 648 | 13.8 | |
| Speed 32.2 - 40.25 km/hr (20 – 25 mph) | | | | |
| Number | Avg. Gap (m) | Number | Avg. Gap (m) | 13.67 |
| 16 | 17.1 | 162 | 19.8 | |
| Speed 40.25 – 48.3 km/hr (25 - 30 mph) | | | | |
| Number | Avg. Gap (m) | Number | Avg. Gap (m) | 12.6 |
| 10 | 19.4 | 78 | 22.2 | |
| Speed 48.3 – 64.4 km/hr ( 30 – 40 mph) | | | | |
| Number | Avg. Gap (m) | Number | Avg. Gap (m) | -4.86 |
| 10 | 27 | 64 | 25.7 | |
| Speed > 64.4 km/hr (40 mph) | | | | |
| Number | Avg. Gap (m) | Number | Avg. Gap (m) | -7.11 |
| 6 | 30.3 | 30 | 28.3 | |

The breakdown of the results has shown that passenger cars do follow heavy vehicles at very close distances at lower speeds, and as the speed increases the following distance in general increases but the percentage at which the passenger cars follow heavy vehicles in comparison to following other passenger cars decreases, while at higher, typically desired speeds, above 48 km/hr (30 mph) for the analyzed car following episodes, passenger cars begin to follow heavy vehicles up to 7% further than they follow other passenger cars.

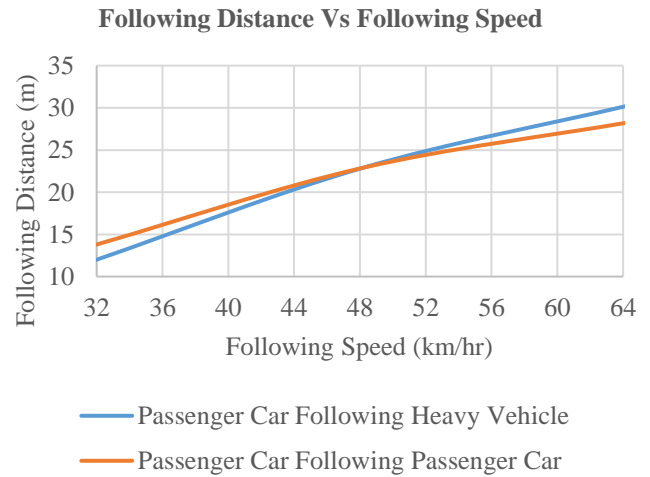

Fig. 4. Change in observed passenger car following distance with change in speed and lead vehicle class.

## B. Passenger Car Lane Change Behavior When Following Heavy Vehicles

From the statistical analysis of the NGSIM dataset, passenger cars were observed keep closer distances to heavy vehicles than that of when following other passenger cars at lower speed conditions, while they were observed to keep a higher gap when speed is well above the typically desired speeds. It was observed that the majority of the passenger car following heavy vehile episodes came to and ending as the following passenger car switched lane. Further statistical analysis was carried out to quantify the lane change behavior of passenger car drivers when following heavy vehicles at the various following speeds observed.

TABLE III. LANE CHANGE BEHAVIOR FOR PASSENGER CARS FOLLOWING HEAVY VEHICLES.

| Following Speed (km/hr) | Lane Change (%) |
|---|---|
| 55.87 (Overall Average) | 33.6 |
| < 20 | 60 |
| 20 - 55 | 42.2 |
| > 55 | 28 |

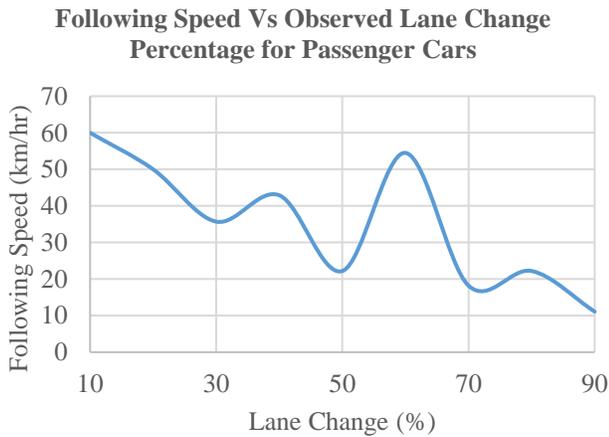

Fig. 5. Change in observed passenger car lane change behavior when following heavy vehicles at various speeds.

The analysis of the lane change behavior of passenger cars indicate high lane change percentage at lower following speed of heavy vehicles, this could be justified as passenger cars aiming to maintain the minimum safe distance while maintaining the maximum possible speed in attempt to find a gap for changing lane to overtake the heavy vehicle ahead in congested traffic conditions, as most of the episodes of passenger cars closely following heavy vehicles were noticed to end when ultimately as the following passenger car changes lane.

Furthermore, it was observed that 40.5% of the passenger cars which have undergone lane change have increased their speed after switching to a new lane and avoiding the heavy vehicle ahead, while 100% of the passenger cars whose initial following speed was below 20 km/hr were observed to increase their speed after changing lanes.

## C. Analysis of Following Behavior Based Against the Predefined GHR Driver Behavior Clusters

To have a better understanding for the car following behavior observed in the statistical analysis of the NGSIM dataset, the car following episodes extracted were further analyzed to assess their compatibility with the different GHR driver's behavioral clusters defined by Higgs and Abbas [1]. The following figures show the count of number of episodes that have conformed with each behavioral cluster based on the class of following and leading vehicles. Each of the car following episodes that lasted a minimum of 25 seconds, obtained from the statistical analysis of the dataset were tested against 30 predefined behavioral clusters based on the class of the following vehicle involved in the car-following episode, and the cluster that produces the minimum root mean squared error was taken as a representative of that specific episode.

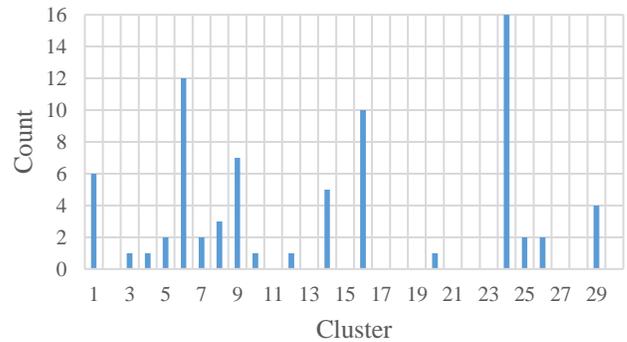

Fig. 6. Cluster Count for Passenger Car / Heavy Vehicle Episodes.

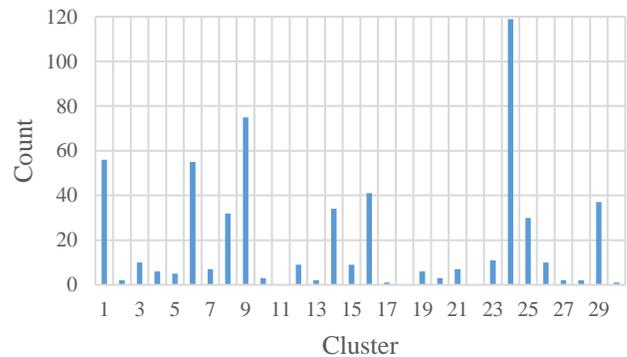

Fig. 7. Cluster Count for Passenger Car / Passenger Car Episodes.

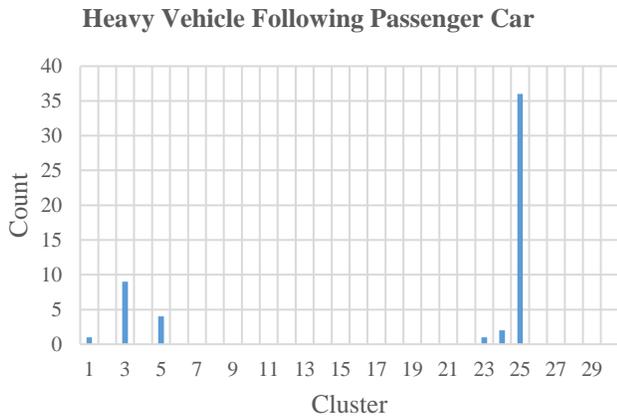

Fig. 8. Cluster Count for Heavy Vehicle / Passenger Car Episodes.

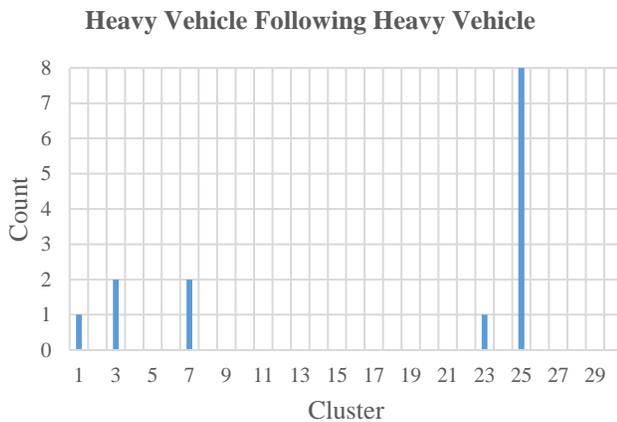

Fig. 9. Cluster Count Heavy Vehicle / Heavy Vehicle Episodes.

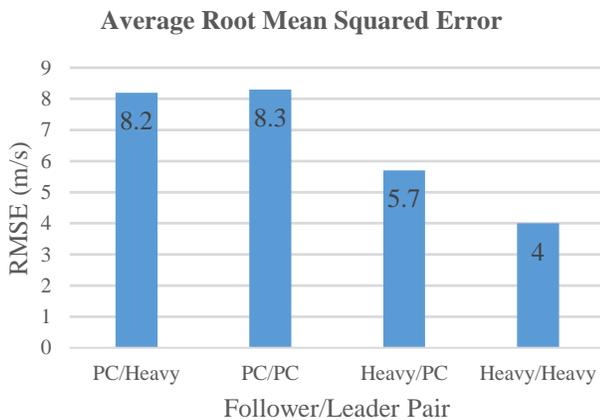

Fig. 10. Average Root Mean Squared Error for each Follower/Leader pair.

The results of the GHR model analysis of the episodes indicate that the passenger car drivers show varying car following patterns, while cluster 24 seemed to be slightly dominant for the passenger car vehicles, most of the other clusters were also represented by an adequate amount of episodes, while the heavy vehicle drivers seem to practice a similar car following pattern as cluster 25 for heavy vehicle drivers represents the majority of episodes examined. Figure 10 shows the frequencies of each cluster observed for different follower/leader vehicle class pairs.

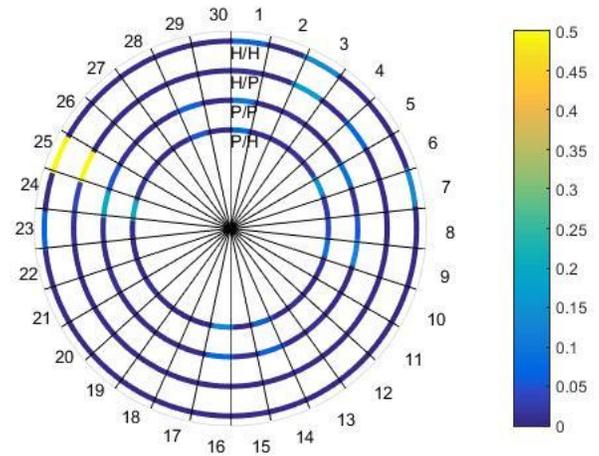

Fig. 11. Circular Plot of Cluster Frequencies for Variant Follower/Leader Pairs.

### D. Studying the Effect of Merge Section on Passenger Car Drivers Following Behavior

The I-80 segment from which the NGSIM trajectories data is obtained has an on-ramp onto the freeway from Powel Street after about 120 meters from the start of the 400 meter monitored segment. The effect of this merge section on the car following behavior was assessed by studying the regression of the GHR model predefined behavioral clusters by Higgs and Abbas [1] at on the following episodes of passenger cars following heavy vehicles at two intervals:

- Beginning of monitored segment until start of the merge.
- Start of merge until end of monitored segment.

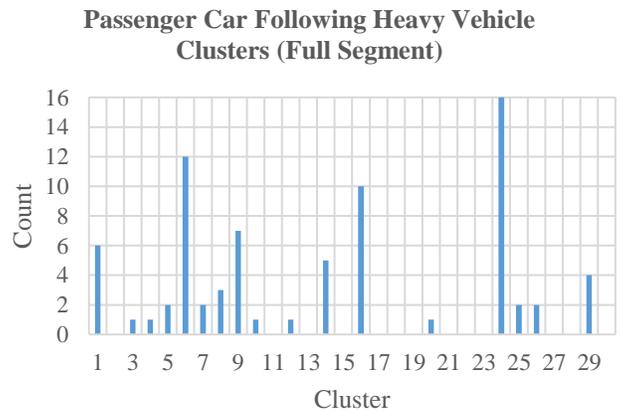

Fig. 12. Cluster Count for Passenger Car / Heavy Vehicle Episodes Over Entire Segment.

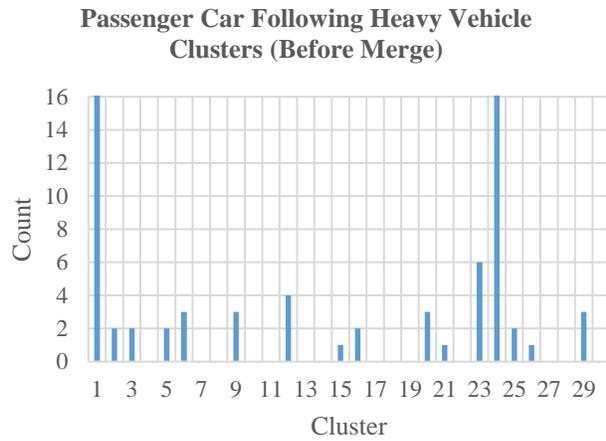

Fig. 13. Cluster Count for Passenger Car / Heavy Vehicle Episodes Before the Merge Section.

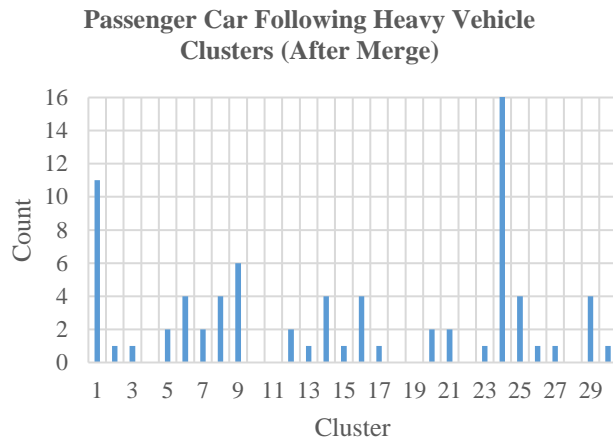

Fig. 14. Cluster Count for Passenger Car / Heavy Vehicle Episodes After the Merge Section.

The results of this analysis indicate that the merge section in the monitored segment has a significant effect on the following behavior observed by passenger car drivers. The passenger car drivers before the merge were mostly represented by clusters 1 and 24. Cluster 24 remained dominant after the merge while cluster 1 significantly decreases after the merge. This could be due to the relatively high perception-reaction time defined for cluster 1 as 2.95 seconds and the drivers are expected to become more alert when entering an area influenced by merging vehicles, which is shown by the perception-reaction time defined for cluster 24 as 1.31 seconds.

Passenger cars before the segment were represented by only 14 clusters, while 23 clusters were needed to represent the passenger car drivers' following behavior after the segment. The average root mean squared error obtained before the merge section is 30% lower than that for after the merge section, which indicates that fewer GHR behavioral clusters could represent the passenger car drivers' following behavior with lower error before the merge while the change in behavior due to the different interactions with the effect of the merge section leads to the use of more behavioral clusters to represent the car following behavior and generates larger errors.

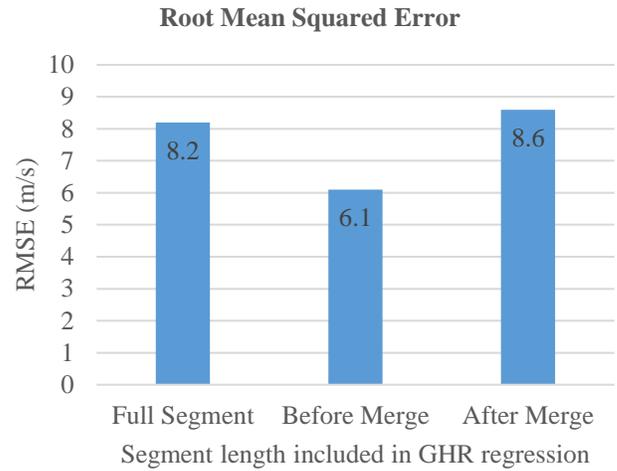

Fig. 16. Average Root Mean Squared Error for Passenger Car / Heavy Vehicle Episodes.

## V. CONCLUSIONS

The results of the statistical analysis in this study validate the results of studies discussed in the literature review, as following distance generally increases as speed increases.

Passenger cars keep closer distances to heavy vehicles at lower speed conditions, while they were observed to keep a higher gap when speed is well above the typically desired speed, with the trade-off point for the analyzed data at speed of about 48 km/hr. As far as the data analyzed, this could be justified as passenger cars trying to maintain minimum safe distance while maintaining the maximum possible speed in attempt to find a gap for changing lane to overtake the heavy vehicle ahead in congested traffic conditions, as most of the episodes of passenger cars closely following heavy vehicles were noticed to end when ultimately as the following passenger car changes lane, while episodes of passenger cars following heavy vehicles at higher speeds were noticed to maintain their following distance and lane throughout the segment, this might also be due to the very short time those vehicles spent in the monitored segment due to their relatively higher speeds and could have possibly undergone lane change to avoid the heavy vehicle ahead further downstream the freeway.

The analysis of the lane changing behavior for passenger cars indicate that lane changing generally seems to be affected by the speed, as it tends to decrease as the speed increases and vice versa. An increase in lane change percentage is observed however between the speeds of 50-60 km/hr, while this could be strictly related to the passenger car drivers' preference, it could also be due to the merge section which forces vehicles traveling at higher speeds to switch lanes to allow merging vehicles to enter the highway.

Significantly greater following distances were observed for the heavy vehicles following each other. This should be studied with more cases and further depth as implementing such changes could prove to be of great significance in traffic streams of high percentage of heavy vehicles.

The application of the GHR clusters obtained by Higgs and Abbas has confirmed that the clustering algorithm does succeed in grouping drivers into defined behavioral patterns. The clusters defined by Higgs and Abbas for the heavy vehicles conform with significantly lower RMSE than that obtained for passenger car driver, this implies that passenger car driver show more varied and discrete following patterns while the heavy vehicle drivers in general tend to drive in similar style, this is emphasized by the outcome obtained for heavy vehicle drivers being significantly represented by a specific cluster.

The study of the effect of the merge section on the GHR clusters used to identify the passenger car's driver behavior when following heavy vehicle show the immense effect that the geometrical configuration of the road section can have on the behavior of the driver, the following behavior of passenger car drivers was observed to significantly be affected by the merge section, while the segmentation of car-following episodes was proven to lead to lower errors in representing the car-following behavior as the drivers may exhibit more than one behavioral cluster during the same car following episode.

The findings of this study show that the car-following behavior is affected by the characteristics and overall speed of the traffic stream and the geometry of the road section, and that the car-following behavior of passenger car drivers is significantly affected by the class of the leading vehicle. The application of GHR model the behavioral clusters implied that the modeling of passenger car's following behavior as a general behavior could be misleading as the car drivers tend to show more varying and discrete behavior which could further vary within a single car-following episode, whereas the heavy vehicle drivers tend to exhibit a profound driving pattern, indicating a similar behavior which could be due to heavy vehicle drivers mainly being professional and trained drivers.

Ultimately, further research is encouraged to develop a better understanding of the car-following process and accordingly develop more accurate and enhanced methods of modeling and simulating traffic that would better represent the stochasticity associated with the car-following process.